\documentstyle[12pt]{article}
\newlength{\abstractwidth}
\begin{document}
\thispagestyle{empty}
\pagestyle{plain}
\def\twog{{\textstyle{{{1}\over{2 g}}}}}
\def\avtwo{{\textstyle{{{9}\over{9+2n}}}}}
\def\avtwov{{\textstyle{{{9}\over{121}}}}}
\def\av{{\textstyle{ {{9}\over{13}} }}}
\def\al{{\textstyle{{{26}\over{11}}}}}
\def\i2{{\textstyle{{{i}\over{2}}}}}
\def\fourg{{\textstyle{{{1}\over{4 g}}}}}
\def\rtwo{{\textstyle{{{2}\over{R}}}}}
\def\twor{{\textstyle{{{1}\over{2R}}}}}
\def\ar{{\textstyle{{{ a l_p^9 }\over{R^3}}}}}
\def\cnn{{\textstyle{ {{ - l (A)^{\delta -1}}\over{b^{(n)} 
(b^{(n)} + \alpha -1)}} \sum_{m\geq 1} {{\delta!}\over{(\delta - m)! m!}}
\sum_{n_i=1}^{n-m} C^{(n_1)} \cdots C^{(n_m)}     }}}
\def\cni{{\textstyle{ {{- l_i (\delta_i)^{n-1} (A_0)^{\delta_i} }} }}}
\def\cn0{{\textstyle{ {{ A_0^{(11/2)}}
\over{11 (1 - A_2) - 26 A_1 A_0 }}}}}
\def\co{{\textstyle{{{24}\over{9}} }}}
\def\lplanck{{\textstyle{l_p}}}
\def\normm{{\textstyle{{{32 \pi^4 n_s}\over{945 (2 \pi \hbar)^9}}}}}
\def\norm1{{\textstyle{{{n_s}\over{945 \cdot 2^4 \pi^5 }}}}}
\def\nr{{\textstyle{ \sqrt{ {{13}\over{44 \cdot a \cdot C }} } }}}
\def\vscale{{\textstyle{{ {{R}\over{l_p}} }}}}
\def\rscale{{\textstyle{{ l_p r }}}}
\def\alp{{\textstyle{{ l_p^3/R }}}}
\def\vint{{\textstyle{ {{|{\bf v}(r) - 
{\bf v}(r')|^4}\over{| {\bf r } - {\bf r'}|^7}} }}}
\def\tension0{{\textstyle{ {{1}\over{{ \sqrt\alpha' } g_s}} }}}
\def\loop{{\textstyle{{{(2 \pi)^2 \lplanck^3}\over{g_s^2}}}}}
\def\tensionp{{\textstyle{(2 \pi)^{-p} (\alpha ')^{-({{p+1}\over{2}})} 
g_s^{-1}}}}
\def\tensionpl{{\textstyle{(2 \pi)^{-p} (\lplanck)^{-(p+1)} 
g_s^{{{(p-2)}\over{3}} } }}}
\def\rel{{\textstyle{\twog \dot{\phi^a_i}\dot{ \phi^a_i} + \i2 \psi^a 
\dot{\psi^a} - \fourg |\phi_i \times \phi_j |^2 + \i2 \epsilon_{abc} \phi^a_i 
\psi^b \gamma^i \psi^c }}} 
\def\s0{{\textstyle{\dot{\X_i}\dot{ X_i} + \i2 \psi \dot{\psi} }}}
\def\half{{\textstyle{1\over2}}}
\def\sq{{\textstyle{{{1}\over{\sqrt 2}}}}}
\def\sqtwo{{\textstyle{{{1}\over{2 \sqrt 2}}}}}
\def\squ{{\textstyle{\sqrt 2}}}
\renewcommand{\thefootnote}{\fnsymbol{footnote}}
\renewcommand{\thanks}[1]{\footnote{#1}} 
\newcommand{\starttext}{
\setcounter{footnote}{0}
\renewcommand{\thefootnote}{\arabic{footnote}}}

\begin{titlepage}
\bigskip
\hskip 3.7in\vbox{\baselineskip12pt
\hbox{EFI--97--27}\hbox{hep-th/9706083}}
\bigskip\bigskip\bigskip\bigskip

\centerline{\large \bf Many--body Dynamics of D0--Branes}

\bigskip\bigskip
\bigskip\bigskip

\centerline{\bf H. Awata\thanks{awata@yukawa.uchicago.edu},
S. Chaudhuri\thanks{schaudhu@yukawa.uchicago.edu}$^1$, M. 
Li\thanks{mli@yukawa.uchicago.edu}, and 
D. Minic\thanks{minic@yukawa.uchicago.edu}$^1$} 
\medskip
\centerline{Enrico Fermi Institute}
\centerline{The University of Chicago}
\centerline{Chicago, IL 60637}

\medskip
\centerline{and}
\medskip

\centerline{$^1$Physics Department}
\centerline{Penn State University}
\centerline{University Park PA 16802}

\bigskip\bigskip

\begin{abstract}
We show that the growth of the size with the number 
of partons holds in a Thomas--Fermi analysis of the 
threshold bound state of D0--branes. Our results
sharpen the evidence that for a fixed value of the
eleven dimensional radius the partonic velocities  
can be made arbitrarily small as one approaches the 
large N limit.
\end{abstract}
\baselineskip=16pt

\end{titlepage}
\starttext
\baselineskip=18pt
\setcounter{footnote}{0}

\section{Introduction}

In this paper we will study the semi--classical behavior 
of a nonrelativistic gas of D0--brane clusters. D0--branes are
point-like solitons carrying the quantum numbers of the first massive
Kaluza--Klein modes of the eleven dimensional supergravity multiplet.
They are conjectured to be the fundamental degrees of freedom of M 
theory in the infinite momentum frame and have a ten dimensional
rest mass $m=\hbar/R$, where $R$ is the radius of the eleventh
dimension. The large $N$ limit of the supersymmetric quantum mechanics
of D0--branes is as yet poorly understood. Our work is a first step
in this direction. The D0--brane clusters are bound by the attractive 
potential arising from their two body interactions forming a threshold 
bound state we will refer to as the \lq\lq D--atom". The universal 
part of this attractive interaction is velocity dependent and its 
leading behavior is of the form $v^4/r^7$. 

For a large atom, the majority of the D--clusters have comparatively 
large quantum numbers so that we can use the semiclassical approximation 
to describe the D--atom. Since our WKB analysis only requires the 
universal interaction between D--clusters we will assume, with
no loss of generality, that they carry no net spin
but will allow for a possible degeneracy of $n_s$ D0--branes per cluster. 
Such a cluster 
is a D--parton carrying $n_s/R$ units of eleven dimensional momentum.
In what follows, we will estimate the growth of the size of the 
D--atom with the total number N of D--partons. The consistency of our 
analysis requires that the de Broglie wavelength of a parton with 
nonrelativistic mass $m$ 
moving in the effective potential $\phi(r)$ of the remaining partons 
must vary only slightly over the characteristic scale of the system 
\cite{landau}, i.e., 
\begin{equation}
{{m \hbar}\over{p^3}} | \phi'(r)| << 1 \quad , \label{semi}
\end{equation}
where $\lambda(r) = (2 \pi \hbar)/ p( r)$ is the de Broglie
wavelength, and  $r$ is the nine dimensional radial coordinate.
The WKB approximation assumes $p^2 \sim \phi $, and
is thus invalid in the vicinity of the turning points 
of the effective potential. Let us scale distances by the eleven 
dimensional 
Planck length $l_p$, with $ r \rightarrow l_p r $. Expressed in 
these units, the WKB condition requires that for a potential of 
the form $\phi \sim r^{- \alpha}$, we have $r^{\alpha -2} << 1$. 

The spatial momentum of the D--atom is equipartitioned among its 
$N$ constituents in the semiclassical approximation. Thus, the D--partons 
populate phase space uniformly with one parton per phase space cell 
filling all available states up to some maximum momentum $p(r)$. This
procedure is independent of the statistics of the D--partons. The 
total number of quantum states in a spatial volume element $dV$ is given by
\begin{equation}
\rho(r)dV = \normm p^9 dV \quad ,
\end{equation}
where $(32 \pi^4/945)|{\bf p}|^{9}$ is the volume of a spherical 
ball in 9d momentum space. 

The growth of the size of the threshold bound state with the number
of D--partons has been estimated in a mean field approximation in \cite{bfss}.
If the atom occupies a box of rescaled size $L^9$ with constant number density 
$\rho_0 $, i.e., $ N= \rho_0 L^9 $, one finds the scaling behavior:  
\begin{equation} 
L \sim N^{1/9}  \quad , \label{naive} 
\end{equation}
which is also in accordance with the holographic bound of one parton
per spatial cell of Planck size area \cite{holo}. We will show that the 
holographic growth of the size of the threshold 
bound state with the number of partons continues to hold when the partonic
interactions are included, at least in the WKB approximation,
thus sharpening the conjecture of Matrix theory. 

Let us use natural units setting $\hbar$$=$$c$$=$$1$. The energy of 
any individual D--parton is $E$$=$$ \half R p^2 - \phi(r)$. Since this 
energy is assumed to be bounded from above we can write down 
a self--consistent relationship between the density and the potential 
within any spatial volume element of the D--atom:
\begin{equation}
\rho(r)=\norm1 (\rtwo )^{9/2} \left ( \phi - \phi_{max} \right )^{9/2} \quad , 
\end{equation}
where $\phi_{max}$ is the value of the potential at the boundary of
the atom which we can set equal to zero. In a conventional 3d atom 
the electrostatic potential $\Phi$ satisfies the Poisson equation. 
Substituting for
the electron number density, we get the Thomas--Fermi equation for 
the self--consistent potential in a large atom \cite{landau}:
\begin{equation}
\Delta \Phi \sim \Phi^{3/2} \quad .
\end{equation}
In the case of the D--atom, the self--consistent potential is dominated
by the velocity dependent gravitational interactions of D--partons. We 
will use the analogous Thomas--Fermi equations for the D--atom to
estimate the growth of its size with the number of D--partons. In contrast
with conventional atoms where $E \sim N^{4/3}$ the nine dimensional 
D--atom displays holographic behavior: the D--parton energy density {\it 
decreases} as the number of partons increases.

\section{The effective potential}

We can infer the general form of the effective potential 
of D--partons from a dimensional analysis of the two--body interactions 
of D0--branes. Neglecting accelerations and time derivatives, the 
relative motion of a pair of nonrelativistic D0--branes in Minkowskian
spacetime is obtained by expanding in powers of the field strength the 
DBI action \cite{witten}\cite{danielsson}\cite{kabat}\cite{douglas}: 
\begin{equation}
S = \tau_0 \int dt ~ {\rm Tr~ det } \left ( \eta_{\mu \nu} + 
2 \pi \alpha ' F_{\mu \nu} \right )^{1/2} + {\rm fermions} \quad , 
\end{equation}
where we have neglected commutator terms for a configuration of
well--separated branes. The D0--brane tension is given by 
$\tau_0=\tension0$ \cite{polchinski}, and we have used the usual 
relations between the Dp--brane tension $\tau_p$,
the radius of the eleventh dimension $R$, and the dimensionless 
string coupling $g_s$:
\begin{equation}
R = g_s^{2/3} \lplanck , \quad \alpha ' = g_s^{-2/3} \lplanck^2 , \quad
\tau_p = \tensionpl \quad . 
\end{equation}
The one--loop approximation to this action 
is given by the supersymmetric matrix quantum mechanics 
of $SU(N)$ matrices, ${\bf \phi}(t)= {\bf \phi}(t)^a \sigma^a $: 
\begin{equation}
S_{0} =  \rel  \quad , \label{relative} 
\end{equation}
reducing to the supersymmetric dynamics of an ordinary coordinate 
world line, $ X(t)= (2 \pi \alpha ') {\bf \phi}_{rel}(t)$, when 
the branes are well--separated as in the semiclassical regime.

By a rescaling of the fermions, the action can be expressed 
as a loop expansion in the dimensionful parameter $g=\loop$ which
plays the role of $\hbar$ in the semiclassical analysis. 
The leading terms 
in the expansion $S= \sum_n g^{(1-n)} S_n$, namely
\begin{equation}
S = g \int dt \left [ (\partial_{t} X^i)^2 + i 
\psi \partial_{t} \psi \right ] +  O(g^0),
\end{equation}
determine the length dimensions of the fields:
\begin{equation}
[X] = -1 , \quad [\partial_{t}] = -1 , \quad [\psi] = -3/2 ,
\end{equation}
consistent with the supersymmetry variations
\begin{equation}
\delta X^{i}=i\epsilon \gamma^{i} \psi , \quad\delta\psi = (\dot{X^i} 
\gamma^i ) \epsilon \quad .
\end{equation}

It is amusing that the velocity dependence of the universal
interaction between D0--branes follows from this dimensional 
analysis when the leading term is a Coulomb potential with 
a $ 1/r^{7}$ fall--off in nine spatial dimensions. Since 
$[S_0]=-4$ the velocity dependence must be of the form $\sim v^4$.
We will set the coefficient in the partonic interaction 
Hamiltonian equal to $a l_p^9/R^3$, where $a$ is an undetermined 
numerical coefficient proportional to the known value for 
ordinary D0--branes \cite{lifschytz}. Thus,
\begin{equation}
\phi (r;{\bf v}) = \ar \vint \quad.
\end{equation} 
The semiclassical estimate for the growth of the size of the
D--atom is independent of the precise value of this coefficient.

\section{The Thomas--Fermi scale} 

We can now write down the nonrelativistic energy density describing
the relative motion of a pair of D--partons with number density and 
velocity $(\rho(r),{\bf v}(r))$ and $(\rho(r'),{\bf v}(r'))$. 
Integrating over the spatial volume gives the Hamiltonian:
\begin{equation}
H = \twor \int v^2(r) \rho (r) dV - \ar \int \rho(r) \vint 
\rho(r') dVdV' \quad.
\end{equation}
The dimensionful parameters in this Hamiltonian can be absorbed
by the rescaling:
\begin{equation}
r \rightarrow \lplanck r , \quad  v \rightarrow \vscale v \quad .
\end{equation}
This scaling renders energy dimensionless and can be motivated by 
the form of the space--time uncertainty relation in string theory
with D--branes \cite{liyoneya}:
\begin{equation}
\Delta X \Delta T \geq \alpha ' \quad , 
\end{equation}
where $\alpha' = \alp $.
The velocity correlations are easily computed in the semiclassical
regime. We have
\begin{equation}
<{\bf v}(r)> = <{\bf v}(r')> = 0 , \quad <{\bf v}^{2n}(r) > = 
\avtwo v^{2n} , 
\quad <({\bf v}(r) \cdot {\bf v }(r'))^2> = \avtwov v^2v'^2 \quad ,
\end{equation}
where $v^2(r)$ is the maximum allowed velocity in the volume element
dV(r). In the semiclassical regime this velocity is self--consistently 
related to the number density through the phase space relation
\begin{equation}
\rho(r) = C l_p^{-9} v(r)^{9}  \quad .
\end{equation}
where $C$$=$$\norm1 $. 
Performing the average over velocities we obtain
\begin{equation}
< |{\bf v}(r) - {\bf v}(r') |^4 > = \av \left ( v^4(r) + v^4(r')
+ \al v^2(r) v^2(r') \right ) \quad .
\end{equation}
Substitution into the Hamiltonian yields velocity dependent nonlocal
interactions. Such interactions can be rewritten in the form of a 
local field theory by introducing three auxiliary scalar fields, 
$\phi_i$, with equations of motion:
\begin{eqnarray}
\Delta \phi_1(r) +  v^9(r) &= 0 \\
\Delta \phi_2(r) + \al  v^{11}(r) &= 0 \\
\Delta \phi_3(r) +  v^{13}(r) &= 0 .
\end{eqnarray}
In simplifying these equations we have rescaled 
$r \rightarrow \nr r$. Varying the Hamiltonian with respect to the 
velocity yields an algebraic equation relating the maximum kinetic 
energy density of the D--partons to the potential energy density 
stored in the three auxiliary potential fields, $U_i = - \phi_i(r)$:
\begin{equation}
11 v^2 = 13 v^4 \phi_1 + 11 v^2 \phi_2 + 9 \phi_3 \quad .
\end{equation}
As boundary conditions on the parton density we require that it 
be normalizable at spatial infinity with a power law fall--off faster 
than a Coulomb potential. Within the bulk of the atom, i.e., at the
origin, the density can be assumed to approach a (positive) constant 
with small power law corrections. We have also found solutions to the 
differential equations with a slow power law decay for which this 
constant vanishes. The detailed analysis of the differential equations 
together with the algebraic constraint equation is given in the appendix.

It is easy to see that partonic interactions will
induce corrections to the naive scaling behavior $L \sim N^{1/9}$
following from the relation $\rho = \rho_0 = N L^{-9}  $.
Consider the power law solutions $v^2 \sim r^{-4/11}$, $\rho = 
r^{-18/11}$.
At distances of order the size of the atom, $r \sim L$, we get
an improved estimate for $\rho$, from which we infer the scaling
$L\sim N^{11/81}$. Restoring the factors of $l_p$, the dimensionful
size grows as
\begin{equation}
L \sim N^{11/81} \lplanck \quad ,
\end{equation}  
for the growth of the size of the D--atom with the number of partons.
Unlike conventional atoms where $N$ plays the role of the
nuclear charge $Z$, and $E \sim Z^{4/3}$, the mean kinetic
energy and number density of the D--atom {\it decrease} upon increasing 
the number of D--partons. We can verify that the estimated atomic size 
lies within the WKB regime. For the above mentioned solution, the 
self--consistent potential $\phi$ scales as $r^{-4/11}$.
At the scale of the atomic size, $L$, using eq.(23) we find the 
scaling behavior $L^{\alpha -2} \sim N^{-2/9}$ 
and the WKB bound is readily satisfied.

The space--time uncertainty relation motivates another analogy 
drawn from atomic physics. Consider the Heisenberg relation for a
highly excited single electron atom. The uncertainty $\Delta X$
scales as $n^2$, the principal quantum number, so that 
$\Delta X \Delta P$$\sim$$ n \hbar$. Let us derive an analogous result
from the uncertainty relations in string theory.
Assuming the scaling behavior $L \sim N^{11/81}$, $v \sim N^{-2/81}$,
with $v \Delta T = \Delta X$, from eq.(15):  :
\begin{equation}
\Delta X \Delta T \geq  l_p^3/R \quad,
\end{equation}
we get the scaling inequality:
\begin{equation}
L^2/v \sim (l_p^3/R) N^{8/27} \quad ,
\end{equation}
with $N$ playing the role of the \lq\lq principal quantum number''. 

\section{Spin dependent forces}

We have neglected spin dependent forces in the Thomas-Fermi analysis.
In this section we will show that in the semi-classical regime where
the Thomas-Fermi approximation applies, all relevant spin-dependent
terms in the two body interaction can indeed be ignored.

As in sect. 2, if we rescale all fields so that there is an overall
coupling $g$ in front of the action, the one-loop term
of the effective Lagrangian for the relative motion of two D0-branes
will be independent of $g$ and will have dimension $-1$. The 
supersymmetric partners of the interaction $v^4/r^7$ will be spin-dependent
interactions at the one-loop level. In addition to the dimensional
analysis, there is another rule that all the terms must observe.
One may assign a quantum number $0$ to $X^i$, $1/2$ to $\psi$,
$-1/2$ to $\epsilon$ and $1$ to $\partial_t$, consistent
with supersymmetry and the form of the action. Thus the quantum 
number $N=N_\partial
+{1/2} N_f$ is the same for the same super-multiplet. This number is
$4$ for $v^4/r^7$. Together with dimensional analysis, this
rule restricts the possible terms which can appear in the 
same supermultiplet as $v^4/r^7$. Schematically, we list the
allowed terms as follows:

$$(\partial_t)^3(\psi^2/r^5),\quad (\partial_t)^2(\psi^4/r^7),
\quad (\partial_t)(\psi^6/r^9), \quad \psi^8/r^{11}.$$
Now, the spin-flip does not occur without interactions, so that for 
the purposes of estimating the contribution of spin dependent 
interactions one can ignore terms containing derivatives of 
fermions. Thus, the relevant terms remaining from the above list are
$$v^3\psi^2/r^8, \quad v^2\psi^4/r^9, \quad v\psi^6/r^{10}, 
\quad \psi^8/r^{11},$$
where the first term can be interpreted as a spin-orbital interaction, 
the second term as a dipole-dipole interaction, the third term as
both, and the last term contains a possible quadrapole-quadrapole
interaction. The magnitude of both the spin-dependent terms and the 
spin-independent term $v^4/r^7$ can be expressed succinctly in
the form $(vr)^n/r^{11}$, where the spin factors can be dropped since
they are independent of $r$ at the
characteristic scale of the large $N$ bound state.
Namely, $(vL)^9\sim \rho L^9\sim N$, so that
$(vL)^n/L^{11}\sim N^{n/9}/L^{11}$. It follows that this
term grows with $n$, the largest value being $n=4$, corresponding 
to the $v^4/r^7$ interaction. Of course, we
could improve upon this estimate by taking into account our previous 
Thomas-Fermi analysis.

\section{Conclusions}

We have shown that the holographic growth of the size of the threshold
bound state with the number of partons \cite{bfss} continues to hold for 
interacting D--partons, at least within the semiclassical approximation.
We have thus sharpened the evidence for one of the key conjectures of Matrix 
theory, namely, that with $R$ fixed as $N \rightarrow \infty$ the 
velocities can be made arbitrarily small at large $N$. We find that the 
qualitative behavior of the mean field analysis survives partonic 
interactions. This raises the interesting question of whether it is 
possible to compute the amplitude for the scattering of two D--atoms within 
the semiclassical regime. This would enable a direct test of 
longitudinal boost invariance within the semiclassical regime which,
while it misses the tail of the distribution, is sensitive to the
bulk of the kinematical phase space explored in the scattering. Boost 
invariance requires physics to only depend on the ratio of the sizes 
of the two bound states. It would be remarkable to understand how the 
holographic growth of particles can avoid contradicting longitudinal 
boost invariance in this picture.

\medskip
\noindent{\bf Acknowledgments}: We would like to thank J. Harvey
and E. Novak for discussions. S.C. and D.M would like to acknowledge 
the warm hospitality of the Enrico Fermi Institute. The work of M.L.
is supported by DOE grant DE-FG02-90ER-40560 and NSF grant PHY 
91-23780. The work of H.A. is supported in part by the Grant--in--Aid
for Scientific Research from the Ministry of Science and Culture, Japan.

\section{Appendix}

We wish to find solutions to the nonlinear differential equation 
\begin{equation}
\Delta \phi = -l \phi^{\delta} \quad , 
\end{equation}
of the form $\phi(r)=A r^a f(r)$ with $\delta$ a positive rational number, 
where $f(r)$ is regular either
at the origin or at infinity. Applying Frobenius' method
we can show that such solutions exist if and only if the leading power
$a$ takes the values $0$, $2-d$, or $- {{2}\over{(\delta -1)}}$. 
Substituting for the radial part of the d--dimensional Laplacian, 
$\Delta_r = r^{1-d} \partial_r (r^{d-1} \partial_r)$, the function 
$f(r)$ can be shown to satisfy the equation:
\begin{equation}
r^2 f''(r) + (2a+d-1) r f'(r) + a(a+d-2) f(r) = -l A^{\delta -1} 
r^{a(\delta -1)+2} (f(r))^{\delta} \quad .
\end{equation}
Let us expand $f(r)$ in the Taylor series 
$f(r)=1+C^{(1)} r^{b^{(1)}} + \cdots $, where the 
$ b^{(i)}$$\geq$$0$ in the limit $r \rightarrow 0$, and 
$b^{(i)} < 0$ in the opposing limit $r \rightarrow \infty$. With 
$a(a+d-2)=0$, and $\alpha$$\equiv$$ 2a+d-1$, we have 
a solution in the form of a recursion relation for the 
coefficients $C^{(n)}$:
\begin{equation}
b^{(n)}=n ( a(\delta -1)+2), \quad C^{(n)}= \cnn  \quad ,
\end{equation}
where $\sum_{i=1}^m n_i = n-1$. 
The boundary condition corresponds to the fields approaching a 
constant in the neighborhood of the origin, i.e., $a=0$. At infinity 
we can have a Coulomb potential with $a=2-d$. In addition, there exist 
solutions to the differential equation with a slow power law decay 
when $a=-2/(\delta -2)$. In nine dimensions such a solution
to the differential equation can have complex coefficients. 

The Thomas--Fermi equations for the D--atom take a similar form,
with $\Delta \phi_i=-l_i \phi_0^{\delta_i}$, and $\phi_0 \equiv v^2$
a positive definite function satisfying the constraint:
\begin{equation}
13 \phi_0^2 \phi_1 + 11 \phi_0 (\phi_2 -1) + 9 \phi_3 = 0 \quad .
\end{equation}
Similarly, we can expand the velocity field in the neighborhood
of the origin: 
\begin{equation}
\phi_0 = A_0 \left ( 1 - \co \cn0 r^2 + \cdots \right )  
\quad ,
\end{equation}
with the relationship between coefficients of the auxiliary 
potentials $ 13 A_1 A_0^2 + 11 (A_2 - 1 )A_0 + 9 A_3 = 0  $.
The velocity distribution can have a 
negative slope in the neighborhood of the origin.

\end{document}